\documentclass[10pt,pre,twocolumn]{revtex4}
\usepackage{graphicx}
\usepackage[floatfix]{epsfig}
\usepackage{amsmath,amssymb,amsthm}
\usepackage{isolatin1}

\begin{document}

\title{Clustering analysis of the ground-state structure of the
vertex-cover problem}

\author{Wolfgang Barthel and Alexander K. Hartmann}

\affiliation{Institut f\"ur Theoretische Physik, Universit\"at
G\"ottingen, Friedrich-Hund-Platz 1, D-37077 G\"ottingen, Germany}

\date{\today}

\begin{abstract}
  Vertex cover is one of the classical NP-complete problems in
theoretical computer science. A vertex cover of a graph is a subset of
vertices such that for each edge at least one of the two endpoints is
contained in the subset. When studied on Erdös-Rényi random graphs
(with connectivity $c$) one observes a threshold behavior: In the
thermodynamic limit the size of the minimal vertex cover is
independent of the specific graph. Recent analytical studies show that
on the phase boundary, for small connectivities $c<e$, the system is
replica symmetric, while for larger connectivities replica symmetry
breaking occurs. This change coincides with a change of the typical
running time of algorithms from polynomial to exponential.
  
  To understand the reasons for this behavior and to compare with the
analytical results, we numerically analyze the structure of the
solution landscape. For this purpose, we have also developed an
algorithm, which allows the calculation of the backbone, without the
need to enumerate all solutions. We study exact solutions found with a
Branch-and-Bound algorithm as well as configurations obtained via a
Monte Carlo simulation.
  
  We analyze the cluster structure of the solution landscape by direct
clustering of the states, by analyzing the eigenvalue spectrum of
correlation matrices and by using a hierarchical clustering method.
All results are compatible with a change at $c=e$. For small
connectivities, the solutions are collected in a finite small number
of clusters, while the number of cluster diverges slowly with system
size for larger connectivities and replica symmetry breaking, but not
1-RSB, occurs.
\end{abstract}

\maketitle

\section{Introduction}
\label{sec:intro}
In combinatorial optimization problems, one has to minimize a certain
function over a discrete phase space consisting of e.g. $2^N$
elements.  Often for a given realization (which we will also call
\emph{instance}) there is more than one point where the function takes
the global minimum value. All these points are called \emph{solutions}
or \emph{ground states}. In our paper we will deal with the phenomenon
of \emph{clustering}: Usually the ground states are not equally
distributed over the phase space. They cluster in one or many groups
that are separated by regions where the function takes values that are
larger than the global minimum.

Such clustering has already been observed in statistical physics when
studying spin glasses \cite{reviewSG}.  For the mean-field Ising spin
glass, also called Sherrington-Kirkpatrick (SK) model \cite{SK},
Parisi has constructed \cite{parisi2}, using the replica trick
\cite{mezard1987,fisher1991}, an analytic solution for the free
energy. This solution exhibits replica-symmetry breaking (RSB), which
means that the state space is organized in an infinitely nested
hierarchy of clusters of states, characterized by ultrametricity
\cite{RaToVi}.  Recently, this solution was mathematically proven to
be the exact one \cite{talagrand2003}.  Also in numerical studies the
clustering structure of the SK model has been observed, e.g. by
calculating the distribution of overlaps \cite{young1983, parisi1993,
billore2003}, when studying the spectrum of spin-spin correlation
matrices \cite{Sinova2000, Sinova2001} or when applying direct
clustering \cite{hed2004}.  For finite-dimensional spin glasses, RSB
seems not to be present fully \cite{KM00,palassini2000} at least not
in the same way as for the mean-field model, since clustering has been
observed numerically but in a different non-ultrametric way
\cite{hed2004}.  On the other hand, for models like Ising ferromagnets
it is clear that they do not exhibit RSB and all solutions are
organized in one cluster.

The use of such analytical tools from statistical mechanics enabled
physicists recently to contribute to the analysis of problems that
originate in theoretical computer science.  Well known problems of
this kind are the satisfiability (SAT) problem \cite{BiMoWe, MeZePa,
MeZe}, number partitioning \cite{Me1, Me2}, graph coloring
\cite{MuPaWeZe}, and vertex cover \cite{WeHa1, HaWe, WeHa2, zhou2003,
cover-review}.  In computer science, one rewrites these optimization
problems as decision problems, i.e. as problems where only the
answers ``yes'' and ``no'' are possible. Here, it means the question
``Is there a solution where the function takes a value less than
$x$''? The above mentioned problems belong to the class $NP$
\cite{GaJo} of decision problems. This means that for any given input
the function can be evaluated easily, i.e. in a time polynomial in
the size of the input (measured e.g. in bits). A single suitable
input, for which the function takes a value less than $x$, proves that
the answer to the decision problem is ``yes''. The open question is
whether one can find a polynomial-time algorithm that for every
possible instance and value of $x$ either constructs such an input or,
in case no such input exists, a proof for the non-existence, which can
be checked in polynomial time.  For the so called {\em NP-complete}
problems up to now only algorithms with an exponentially growing
running time in the worst case are known. But these instances are in
some regions of the instance space exponentially rare so one can do
better in the \emph{typical} case.

What is ``typical'' cannot be defined uniquely, hence one has to study
suitable parametrized (usually random) ensembles of problem instances.
By varying these parameters one often finds phase boundaries which
separate regions where the answer to the decision problem is ``yes''
resp. ``no'' with probability one \cite{review1,review2}.
Analytically, the phase diagrams of these problems can be studied
using some well-know techniques from statistical physics, like the
replica trick \cite{monasson1996,nature}, or the cavity approach
\cite{MeZe}. But full solutions have not been found in the most cases,
since the problems from theoretical computer science are usually not
defined on complete graphs but on diluted graphs, which poses
additional technical problems. Usually, one can only calculate the
solution in the case of replica symmetry
\cite{monasson1996,nature,BiMoWe}, or in the case of one-step replica
symmetry breaking (1-RSB) \cite{MeZePa,MeZe}, and look for the
stability of the solutions.  For this reason, the relation between the
solution and the clustering structure is not well established and it
is far from being clear for most models how the clustering structure
looks like.  However, most statistical physicist believe that the
failure of replica symmetry (RS) leads indeed to clustering
\cite{weigt2004,zecchina2004}.  So far, only few analytical studies of
the clustering properties of classical combinatorial optimization
problems like SAT have been performed \cite{BiMoWe,mezard2003}. These
results depend or may depend on the specific assumptions one makes
when applying certain analytical tools and when performing
approximations.  In particular, it is unlikely that the clustering of
models on dilute graphs is exactly the same as it is found for the
mean-field SK spin glass.  So from the physicist point of view, it is
quite interesting to study the organization of the phase space using
numerical methods to understand better the meaning of ``complex
organization of phase space'' for other, non-mean-field models, like
combinatorial optimization problems.  It is the aim of this paper, to
study numerically the clustering properties of one particular problem,
the vertex-cover problem (see below) using three different
complementary approaches.

The study of the solution structure is not only important for
physicists, but also of interest for computer science.  From an
algorithmic point of view, especially the ground-state structure seems
to play an important role.  If it consists only of a single cluster,
finding a solution typically will be easy. In this case one can often
construct algorithms that quickly detect the promising regions of the
solution space. On the other hand the appearance of clustered ground
states is often accompanied by the existence of suboptimal local
minima of the function to be optimized \cite{weigt2004}. They mislead
local algorithms and make computation expensive. In this case, the
typical computation time grows exponentially in the the size of the
instance.

Such \emph{easy-hard} transitions have been observed in many
optimization problems, first by studying SAT numerically
\cite{mitchell1992,selman1994}.  For SAT, the ``yes''-phase (which is
referred to as SAT-phase) is split up into two regions: an
\emph{easy}-SAT and a \emph{hard}-SAT phase, which have exactly the
properties described above. For all known algorithms the onset of
exponential median running times is in the \emph{easy} phase, although
during the last years better and better heuristics extended the region
of instances that can be solved typically in polynomial time. However
it is an open question, whether this phase boundary really gives an
upper bound for the best heuristics possible.

In our paper we deal with the minimal vertex-cover (VC) problem. We
consider random graphs $G=(V,E)$ with $N$ vertices $i \in {1, 2, \dots
N}$ and $\frac c 2 N$ randomly drawn, undirected edges
$\left\{i,j\right\} \in E \subset V \times V$, each connecting a pair
of vertices. In this notation $c$ is the connectivity, i.e. the
average number of edges each vertex is contained in.

\label{sec:proprandomgraphs}
Let us briefly recall properties of random graphs that are relevant to
our analysis of the ground-state structure \cite{ErRe}.  For $c<1$ the
typical random graph only consists of small trees each with size of
${\cal O}(1)$. Additionally there is a finite \emph{number} of
components, also with size of ${\cal O}(1)$, each having one closed
loop, e.g. for $N\rightarrow \infty$ the fraction of closed loops
approaches zero \cite{Bo}.  For $c>1$, the finite-size tree-like
components and the components with loops remain, but there is one
additional component which contains ${\cal O}(N)$ vertices, the
so-called \emph{giant component}.

Let $V' \subset V$ be a subset of all vertices. We call a vertex $v$
\emph{covered} if $v\in V'$, \emph{uncovered} if $v\notin V'$.
Similarly an edge is \emph{covered} if at least one of its endpoints
is covered.  If all edges of $G$ are covered, then we call $V'$ a
vertex cover $V_{VC}$.  We denote $X\equiv|V'|$ and $x\equiv X/N$.

For a graph $G=(V,E)$ the minimal VC problem is the following
optimization problem: Construct a vertex cover $V_{VC-min}\subset V$
of minimal cardinality and find its size $X_{\min}\equiv
\left|V_{VC-min}\right|$.  Usually there are many solutions of the
same size. The {\em backbone} consists of those vertices, which appear
in all solutions in the same manner, i.e. which are always covered
or always uncovered.

Algorithmically, one can solve the minimal vertex-cover problem
independently for each component of the underlying graph.  Any
combination of the vertex covers of the individual components gives a
VC for $G$.  For $c<1$, where no giant component exists, since the
different components are independent and of size ${\cal O}(1)$, we
cannot expect a complicated ground-state structure. Furthermore, as we
show in the appendix, the solution structure for trees is always
simple.  Hence the main emphasis of the paper will be on studying the
giant component appearing for $c>1$, since only this component can be
responsible for a complex ground-state structure.

The vertex-cover problem on random graphs exhibits the threshold
phenomenon described above.  For minimum covers, in the limit
$N\rightarrow \infty$, one expects that the fraction $x_{\min}$ of
covered vertices depends only on the connectivity $c$, i.e. we have
$x_{\min}=x_{\min}(c)$.  For $x<x_{\min}(c)$ almost no graphs with
connectivity $c$ have a VC of this size, on the opposite for $x\ge
x_{\min}(c)$ such a cover can be found with probability 1.

By applying the replica method \cite{mezard1987,fisher1991} one can
derive analytical results \cite{WeHa1} for this phase boundary.  In
the language of statistical physics, we can think of the size of a VC
as its {energy}, and of VCs of minimal size as {ground states}.  Using
a replica symmetric (RS) ansatz one gets $x(c) =
1-\frac{2W(c)+W(c)^2}{2c}$, where $W(c)$ is the Lambert-$W$-function
given by $c = W(c)\exp({W(c)})$. By studying the stability of the
solution and by comparison with numerical results, one finds that the
RS solution is valid in the region $c<e$ (where $e\approx 2,718$ is
the Euler number). This has also been proven rigorously
\cite{bauer2001a} by analyzing a specific algorithm, the leaf removal
algorithm \cite{bauer2001} which we will describe in section
\ref{sec:algos}.
 
The RS ansatz assumes that all ground states form a single
cluster. This assumption seems to be violated for $c>e$, where one can
construct analytically solutions with smaller fraction $x$ of covered
vertices. One can extend the calculation by including RSB, here we
expect that the ground states form clusters that are separated by
extensive energy barriers.  A single level of clustering corresponds
to one-step replica symmetric breaking (1-RSB). If the clusters itself
have some hierarchical clustering structure, i.e. a set of very
similar solutions is subdivided in a structured manner in subsets of
even more similar solutions, $n$-step-RSB or full-RSB (the last being
the case where $n = \infty$) appears. However, this full-RSB is not
necessarily the same as found in the SK model.

In our paper we will analyze the ground-state structure of VC
numerically. Especially we will focus on the behavior of the cluster
structure around $c=e$. We have studied different definitions of
clusters and methods to detect nontrivial clustering.  They have in
common that solutions which are very similar to each other are
considered to be in one cluster. Definitions and details are given
later on.  So far it is not clear to what extend the observation of
clustering phenomena depends on the definition of the clusters applied
and which one is the ``correct'' one to describe RS or RSB.
Nevertheless, the results presented below for the different methods
turn out to be compatible with onset of clustering at $c=e$,
supporting the previous analytical findings.

The rest of the paper is organized as follows: First we will describe
in detail algorithms that we used to find minimal VCs:
Branch-and-Bound, our backbone algorithm and a Monte Carlo approach.
In the main section \ref{sec:clustering} we will present three
different methods for analyzing the ground-state structure and the
corresponding results: direct clustering, analysis of the eigenvalue
spectrum of correlation matrices and Ward's algorithm, which is a
hierarchical clustering method.  Finally, we will summarize and give
an outlook.


\section{Algorithms for finding ground states}
\label{sec:algos}

The vertex-cover problem is NP-complete, so all known algorithms have
a solution time that in the worst case grows exponentially with the
number of variables. In the typical case, algorithms often perform
better. This behavior seems to be closely related to the cluster
structures, which we study in this work.  For connectivities $c<e$, a
minimal VC can be found typically in a time polynomial in the number
of vertices using the leaf removal algorithm, which is explained
below.  Thus the problem is typically easy in this region of
connectivity. On the contrary there is no similar algorithm known for
random graphs with $c>e$. So the point $c=e$ is also interesting from
an algorithmic point of view.

We use two different methods to generate VCs. Both will be explained
in this section:

\begin{enumerate}
\item exact enumeration of all ground states
\item sampling the structure of close-to-minimum covers with Monte
Carlo methods
\end{enumerate}

\subsection{Exact enumeration}
\label{sec:exactenumeration}

The ideal case to study the cluster structure of the solutions is to
have all solutions available. Since the number of solutions grows very
fast with system size $N$, a direct enumeration is not the best
choice. We will now explain in several steps the algorithms we have
used.  First we always split up the graph into its connected
components, because they can be treated independently.

We now explain, how {\em one solution} can be found (for each
component). As a first step, we apply the \emph{leaf-removal
algorithm} \cite{bauer2001}, which is a special variant of the
algorithm by Tarjan and Trojanowski \cite{tarjan77}.  The idea of leaf
removal is the following: A leaf of the graph is a vertex $i$ with
connectivity one, i.e. it has only one neighbor vertex $j$. In a VC
either $i$ or $j$ has to be covered. If we covered the leaf $i$ then
only the edge between $i$ and $j$ would be covered. Thus we can cover
$j$ and so all edges originating from $j$ are covered including the
one to $i$. We no more need to consider these edges, therefore we
remove them from the graph, possibly creating new leaves. We
iteratively repeat this procedure until the graph is empty or no more
leaves are present and the so-called \emph{core} remains. These steps
take polynomial time in $N$.  Bauer and Golinelli show that for $c\le
e$ this core is composed of small components of size $\log(N)$, while
for larger $c$ a complex structure of size $O(N)$ remains.

The core has to be treated with a more elaborate method, the
Branch-and-Bound algorithm. Its basic idea is that all possible
configurations can be represented as a binary tree. At each node the
tree splits into two subtrees corresponding to setting one vertex to
covered or to uncovered. Some of the $2^N$ leaves correspond to vertex
covers, some even to minimal vertex covers. The algorithm starts at
the root node, by selecting any vertex. The order the vertices are
selected is in principle arbitrary. A good performance can be
obtained, when e.g. selecting the vertices in the order of the current
degree, i.e. the number of currently uncovered neighbors.  Since at
this point we do not know which vertices have to be covered, we have
to \emph{branch}: We set one of the variables to one of the two
possible values, i.e. we go down one of the branches that start at
the root node. Iteratively we continue this procedure until a full VC
has been found.  Then we go back to an earlier branching point, one
calls this {\em backtracking}, and explore the other subtrees.

Note, after setting a vertex to covered, new leaves may appear in the
graph.  In this case, we remove them by applying leaf removal
again. The removed vertices have to be inserted again, when
backtracking.  Furthermore, since we are interested only in full
covers, we can always cover all neighbors of a vertex we have
uncovered.

A significant speed-up can be achieved by \emph{bounding}.  The basic
idea is to omit subtrees, where for sure no minimum vertex cover can
be found. This can be achieved, by keeping track of of the smallest
cover $X_{\min}$ found so far. Let now, at any stage when building the
tree, $X<X_{\min}$ be the current number of covered vertices and let
$d(j)$ be the current number of uncovered edges that connect to the
uncovered vertices $j$. Since we are looking for a smaller cover than
$X_{\min}$, we want to cover at most $k=X_{\min}-X-1$ additional
vertices. By covering $k$ vertices we can reduce the number of
uncovered edges at most by $M = \max_{j_1,\dots,j_k}d(j_1)+\dots
d(j_k)$. If the number of currently uncovered edges is greater than
$M$ we can omit this branch of the search tree since it cannot lead to
a new optimum. Note that the Branch-and-Bound algorithm takes in the
worst case an exponential running time. Since the core is only of
order ${\cal O}(\log N)$ for $c<e$, this results in a polynomial
running time in combination with leaf-removal for these values of $c$.
 
Having found a single ground state, in order to enumerate all
solutions, we can now again reduce the size of the problem by
identifying the vertices that have in all minimum solutions the same
state, the so-called \emph{backbone}. We first consider the covered
backbone, i.e. vertices which are covered in all minimum solutions.

\begin{figure}[htbp]
  \centering \includegraphics[width = 8cm]{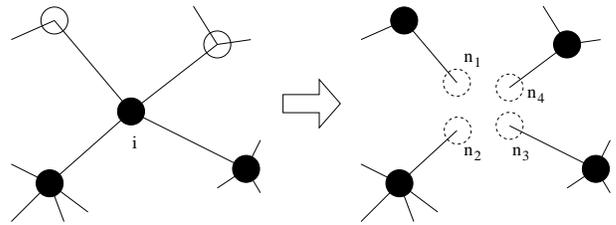}
  \caption{Identifying the backbone: A vertex $i$ is fixed to be
    uncovered by replacing it with new vertices, one for each edge
(note: a leaf of the original graph can never belong to the covered
backbone). $i$ belongs to the covered backbone, iff the minimal VCs of
the new graph are larger than in the original one.}
  \label{fig:identbb}
\end{figure}

Suppose that these solutions have $X$ vertices covered and vertex $i$
is in the covered backbone. If we fixed $i$ to be uncovered then we
could only find vertex covers with at least $X+1$ covered
vertices. This is the idea of our backbone algorithm (cf. Fig.\
\ref{fig:identbb}):

\begin{enumerate}
\item Select a covered vertex $i$
\item For each edge $(i,j)$ add a new vertex $n_j$ and a new edge
$(n_j,j)$ to the graph
\item Remove vertex $i$ and all its edges from the graph
\item Find the ground-state energy (cover size) of the new graph
$G'$. Since all vertices $n_j$ are now leaves, $G'$ has a ground state
with all $n_j$ uncovered. If $X(G') = X(G)$ then we also have found a
ground state of $G$, with vertex $i$ uncovered. Obviously the converse
is also true. So we have $X(G') = X(G) \Leftrightarrow i \notin
\text{covered backbone}$
\end{enumerate}

\begin{figure}[htbp]
  \centering \includegraphics[width = 8cm]{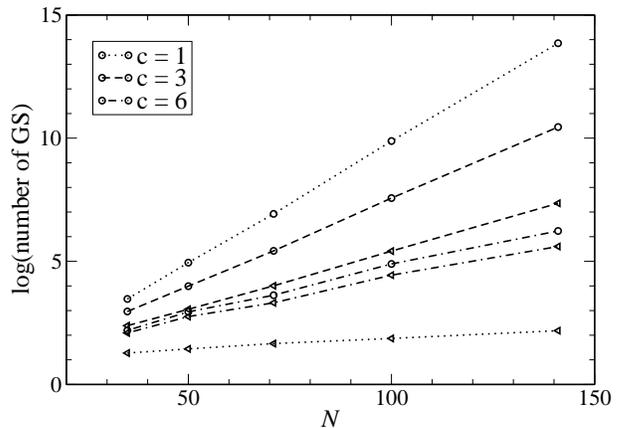}
  \caption{Median number of ground states of the largest component
    for different values of $c$. The circles represent values before
removal of the backbone, the triangles after the removal (error bars
are at most of symbol size). The smaller $c$ the larger is the
speedup, which is due to reduced size of the largest component. }
  \label{fig:bbgs}
\end{figure}

Since the backbone vertices are fixed and all adjacent edges are
covered, we can remove them from the graph without changing the number
of solutions. Vertices, that have only neighbors belonging to the
covered backbone are uncovered in all solutions, they form the
\emph{uncovered backbone}. After the removal of the covered backbone
they become isolated vertices and can be removed as well. Since the
backbone size is rather large \cite{cover-review}, the remaining graph
often breaks apart into different components which can be treated
individually. This speeds up the Branch-and-Bound algorithm when we
now enumerate all ground states. In Fig.\ \ref{fig:bbgs} we compare
the median number of ground states of the largest component before and
after backbone removal, respectively.  In both cases the number grows
exponentially, but the exponent is reduced, especially for smaller
$c$. This can be easily understood, since e.g. a single component of
two vertices that gets separated from the largest component due to the
removal of the backbone reduces the number of ground states of this
component by a factor of two.

For the enumeration of all ground states, we use a variant of the
Branch-and-Bound algorithm without leaf-removal. Also we allow at each
node for $k=X_{\min}-X$ additional covered vertices instead of
$k=X_{\min}-X-1$, where $X_{min}$ is the size of a minimal vertex
cover for the current component, which is known from the first step.
We store all covers which have the size $X_{\min}$. When the algorithm
terminates, all minimum covers are stored.

To summarize, the general outline of the algorithm is as follows:

\begin{tabbing}
xxx\=xxx\=xxx\=xxx\=xxx\=xxx\=xxx\=xxx\=\kill
\noindent{\bf begin}\\
\>split the graph into connected components;\\ \>for each component\\
\>{\bf begin}\\ \>\>find \emph{a single solution} using leaf removal\\
\>\>\> and Branch-and-Bound;\\ \>\>determine the backbone vertices;\\
\>\>remove all backbone vertices;\\ \>\>split the graph into connected
components;\\ \>\>for each component\\ \>\>{\bf begin}\\
\>\>\>enumerate \emph{all solutions} using\\ \>\>\>\>
Branch-and-Bound;\\ \>\>{\bf end}\\ \>{\bf end}\\ {\bf end}
\end{tabbing}

The complete set of all solutions contains all possible combinations
of covers for all components. Since the trees contribute only a
trivial background to the solution landscape, we analyze in the
following very often only the largest component of each graph.

The complete algorithm can find {\em one} minimal VC for graphs with
sizes between $N \approx 200$ (for $c=6$) and $N \approx 2000$ (for
$c=3$). The number of solutions grows strongly exponential, so
enumeration of all ground states is possible only up to $N \approx
100$.

\subsection{Monte Carlo algorithm}
\label{sec:ptalgo}

  For larger graphs, we apply a parallel tempering (PT)
\cite{marinari1992,hukushima1996} Monte Carlo (MC) \cite{landau2000}
algorithm to sample configurations.

The basic approach is that we simulate the behavior of an equivalent
system, the \emph{hard-core lattice gas} \cite{WeHa2}. The graph
corresponds to a lattice with edges of length one. Each vertex
corresponds to a site of the lattice that can be occupied by a hard
sphere with radius one. The states \emph{covered/uncovered} of the
vertices correspond to the two possibilities \emph{not
occupied/occupied} in this order.  Hence, for a given cover $U$ of the
graph, we assign an \emph{occupation number} $x_i$ to each site of the
lattice:
 \begin{equation}
   \label{eq:occnum}
   x_i :=
   \begin{cases}
     0 &\text{if corresponding vertex } i \in U\\ 1 &\text{if
corresponding vertex } i \notin U
   \end{cases}
 \end{equation}

The condition, that in a VC for each edge at least one of its vertices
has to be covered, implies for edges on the lattice that at most one
of the two endpoints can have the occupation number $1$. In other
words, if a sphere is put on site $i$ then all sites that are
connected to $i$ by an edge cannot be occupied. A given assignment to
the occupation numbers is thus a VC, if the characteristic function
\begin{equation}
  \label{eq:char_func_hslg}
  \chi(\vec{x}) = \prod_{ \{i,j\} \in E} (1-x_i x_j)
\end{equation}
equals one. We can control the number of spheres by applying the
grand-canonical formalism. The grand-canonical partition function
$\Xi$ is given by
\begin{equation}
  \label{eq:part_hslg}
 \Xi = \sum_{{x_i = 0,1}} \exp\left(\mu \sum x_i\right)\chi(\vec{x})
\end{equation}
where $\mu$ is the chemical potential. Configurations with a larger
number of spheres get an exponential greater weight. In the large
$\mu$-limit the sum is dominated by the configurations where the
largest number of hard spheres is put on the lattice. These
configurations correspond to minimal VCs.

\begin{table}[t]
\begin{tabbing}
xxx\=xxx\=xxx\=xxx\=xxx\=xxx\=xxx\=xxx\=xxx\=xxx\=xxx\=xxx\=\kill {\bf
begin}\\ \>initialize configuration $\left\{x_i^{(1)}\right\} \dots
\left\{x_i^{(n)}\right\}$ randomly\\ \>{\bf for} $t=1 \ldots N_{\rm
MC}$ {\bf do}\\ \>{\bf begin}\\ \>\> {\bf for} each copy $k=1\ldots n$
{\bf do}\\ \>\>\> {\bf do} $N$ times\\ \>\>\>{\bf begin} \{perform one
MC step\}\\ \>\>\>\> choose a random vertex $v$\\ \>\>\>\> with
probability 1/2 do step (M) or (E)\\ \>\>\>\>(M)\> {\bf if} $v$ is
covered and has exactly one\\ \>\>\>\>\> uncovered neighbor $w$ {\bf
then}\\ \>\>\>\>\>\> uncover $v$ and cover $w$\\ \>\>\>\>\> {\bf else}
do nothing\\ \>\>\>\>(E)\> {\bf if} $v$ is uncovered {\bf then}\\
\>\>\>\>\>\> cover it with prop. $e^{-\mu_i}$\\ \>\>\>\>\> {\bf else
if} $v$ and all its neighbors \\ \>\>\>\>\> are covered {\bf then}\\
\>\>\>\>\>\> uncover $v$\\ \>\>\>{\bf end}\\ \>\> {\bf for} $k=1\dots
n-1$\\ \>\>{\bf begin} \{perform PT moves\}\\ \>\>\> set $\Delta E_k =
(\mu_k - \mu_{k+1})$ $\cdot ( \sum_i x_i^{(k)} - \sum_i
x_i^{(k+1)})$\\ \>\>\> with prop. $\exp(-\min(\Delta E_i,0))$\\
\>\>\>\> exchange $\left\{x_i^{(k)}\right\} \leftrightarrow
\left\{x_i^{(k+1)}\right\}$\\ \>\>{\bf end}\\ \>{\bf end}\\ {\bf end}
\end{tabbing}
\caption{Parallel tempering Monte Carlo algorithm. $N_{\rm MC}$
  denotes the number of MC sweeps per copy, $n$ the number of
different copies}
\label{fig:alg}
\end{table}

The MC moves \cite{WeHa2} which sample Eq. (\ref{eq:part_hslg})
consist in selecting a vertex randomly and performing with probability
$p=0.5$ either a move (M) or an exchange (E) step. For details, see
the algorithm in Tab.\ \ref{fig:alg}.

The MC simulation is performed within a PT \cite{marinari1998}
framework. Its idea is that different copies of the system (for the
same graph) are simulated each at a different value of the chemical
potential $\mu$. At lower values of $\mu$ spheres can be more easily
removed than at higher ones, so the system can overcome larger energy
barriers. At high values of $\mu$ the system equilibrates to a local
minimum. The basic PT step is to perform exchanges of the
configuration for neighboring values of the chemical potential in a
way such that globally detailed balance is ensured, for details, see
Tab.\ \ref{fig:alg}

\begin{figure}[htbp]
  \centering \includegraphics[width = 8cm]{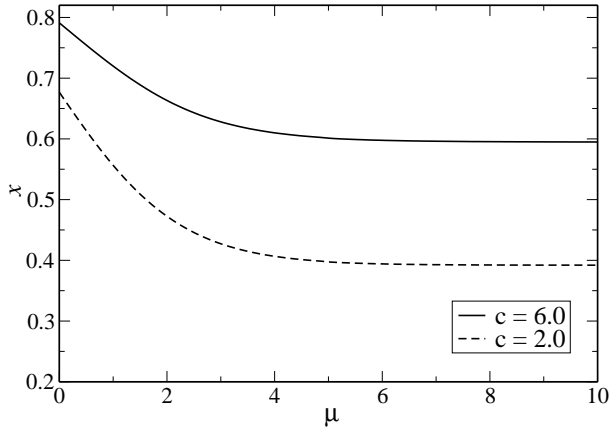}
  \caption{Average size $x$ of the vertex cover for different values of
    the chemical potential $\mu$ (N=64000). In the large-$\mu$ limit
$x$ approaches the average size $x_c$ of the minimal vertex cover}
  \label{fig:rhomu}
\end{figure}

In our simulation we use $n=26$ values between $\mu=1$ and $\mu=12$.
The simulations run for $N_{\rm MC}=10^6$ MC-sweeps.  $500$
configurations are saved for a value $\mu = 9$. At this value of $\mu$
most of the sampled states have the ground state energy, cf. Fig.
\ref{fig:rhomu}. For the system sizes that are tractable by the exact
algorithm (cf. beginning of subsection) we additionally verified that
parallel tempering really finds ground states.

\section{Clustering methods}
\label{sec:clustering}
\subsection{Clustering using hamming distance}

\label{sec:hamdist}
Our first naive approach for analyzing the ground-state structure is
based on the hamming distance between different solutions. The hamming
distance
$dist_{ham}(\left\{{\vec{x}}^{(\alpha)}\right\},\left\{{\vec{x}}^{(\beta)}\right\})
\equiv d_{\alpha\beta}$ of two solutions is the number vertices in
which the two configurations differ. If for two optimal solutions
their hamming distance is minimal, i.e.
$dist_{ham}(\left\{{\vec{x}}^{(\alpha)}\right\},\left\{{\vec{x}}^{(\beta)}\right\})
= 2$, we will call them \emph{neighbors}.  Since for a given
realization all ground states have the same energy, neighboring states
differ only in two vertices $i$ and $j$ which are linked by an edge.
In other words, one can get a {neighboring} state
${\vec{x}}^{(\alpha)}$ of a given ground state by choosing a covered
vertex $i$ which has the property that all but one vertex $j$ of the
adjacent vertices of $i$ are covered. The state with $i$ uncovered and
$j$ covered is a neighboring ground state of
${\vec{x}}^{(\alpha)}$. If we think of \emph{covering marks} put on
each covered vertex, then this is equivalent to moving a covering mark
along an edge to an adjacent vertex. Step (M) of the MC algorithm
above exactly corresponds to this move.

We define a \emph{cluster} $\cal C$ as maximal set of ground states,
that can be reached by repeatedly applying the above procedure.
Similar definitions of clusters have been used e.g. for the analysis
of random p-XOR-SAT \cite{mezard2003} or finite-dimensional spin
glasses \cite{alex-valleys-long}.  States which belong to different
clusters are separated by a hamming distance of at least 4. In
appendix \ref{app:prooftree} we will show that for a graph, which is a
tree, the ground-state structure always consists of exactly one single
cluster.

To decide that two arbitrary ground states ${\vec{x}}^{(\alpha)}$ and
${\vec{x}}^{(\beta)}$ do not belong to the same cluster, one needs to
calculate the complete cluster ${\vec{x}}^{(\alpha)}$ (or
${\vec{x}}^{(\beta)}$) belongs to. Hence the clustering is very
expensive.

The naive algorithm is as follows:
\begin{enumerate}
\item identify the giant component of the graph (we ignore the $O(1)$
components since they do not influence the cluster structure,
cf. Sec.\ \ref{sec:proprandomgraphs})
\item calculate all ground states ${\vec{x}}^{(\alpha)}$ as described
in section \ref{sec:algos}
\item cluster the ground-state configurations:
\begin{tabbing}
xxx\=xxx\=xxx\=xxx\=xxx\=xxx\=xxx\=xxx\=\kill {\bf begin}\\ \>$S$:=
set of all ground states\\ \>$i = 0$ \{number of so far detected
clusters\}\\ \>{\bf while} $S$ not empty {\bf do}\\ \> {\bf begin}\\
\>\>$i = i + 1$\\ \>\>remove an element ${\vec{x}}^{(\alpha)}$ from
$S$\\ \>\>set cluster $C_i=({\vec{x}}^{(\alpha)})$\\ \>\>set pointer
${\vec{x}}^{(\beta)}$ to first element of $C_i$\\ \>\>{\bf while
${\vec{x}}^{(\beta)} <> NULL$} {\bf do}\\ \>\> {\bf begin}\\
\>\>\>{\bf for all} elements ${\vec{x}}^{(\gamma)}$ of $S$\\
\>\>\>\>{\bf if} $d_{ham}({\vec{x}}^{(\beta)},
{\vec{x}}^{(\gamma)})=2$ {\bf then}\\ \>\>\>\>{\bf begin}\\
\>\>\>\>\>remove ${\vec{x}}^{(\gamma)}$ from $S$\\ \>\>\>\>\>put
${\vec{x}}^{(\gamma)}$ at the end of $C_i$\\ \>\>\>\>{\bf end}\\
\>\>\>set pointer ${\vec{x}}^{(\beta)}$ to next element of $C_i$\\
\>\>\> or to $NULL$ if there is no more\\ \>\>{\bf end}\\ \>{\bf
end}\\ {\bf end}
\end{tabbing}

\end{enumerate}

\begin{figure}[htbp]
  \centering \includegraphics[width = 8cm,clip]{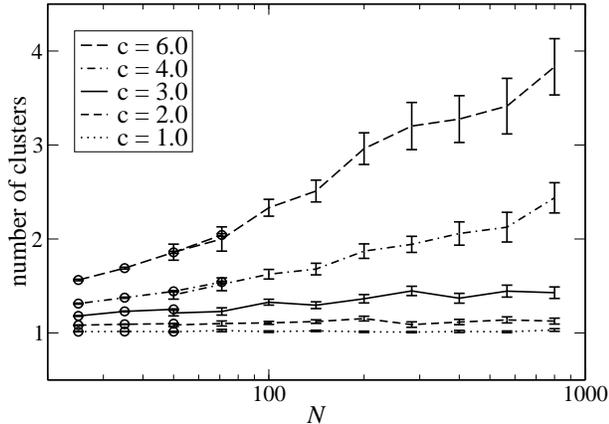}
  \caption{Average number of clusters in the solution space of the largest
    component as function of system size. The circle symbols for small
system sizes have been obtained by clustering complete sets of ground
states. For large systems we sampled ground states with a Monte Carlo
algorithm at large but finite chemical potential $\mu$.}
  \label{fig:clusterneighbour}
\end{figure}

 The crucial point is that one really needs to consider all ground
states and not just a sample. The algorithm is quadratic in the number
of ground states ${\vec{x}}^{(\alpha)}$, which makes the method
applicable to system sizes up to $N \approx 70$, depending on the
connectivity $c$.  For every value of $N$ we sampled $10^4$
realizations. The average number of clusters as function of
connectivity is shown as circles in figure
\ref{fig:clusterneighbour}. We mainly use this naive method to judge
the validity of its extension which will be described in the next
section \ref{sec:cluster_path}. For $c<e$ the number of clusters
remains close to one. For larger values of $c$ the number of clusters
increases with system size. Apparently the increase is compatible with
a logarithmic growth as a function of system size, see discussion in
the next section.

\subsection{Cluster detection using sampled states}
\label{sec:cluster_path}

In this section we will show, how one can identify clusters when only
a small fraction of the solution space has been sampled, as obtained
by using Monte Carlo methods such as parallel tempering. We start in
one of the configurations ${\vec{x}}^{(\alpha)}$ and follow a local
exchange dynamics which does not change the energy, i.e. the size of
the cover. If we can reach another configuration ${\vec{x}}^{(\beta)}$
then ${\vec{x}}^{(\alpha)}$ and ${\vec{x}}^{(\beta)}$ are in the same
cluster, according to the cluster definition above.

Let us compare two ground states ${\vec{x}}^{(\alpha)}$ and
${\vec{x}}^{(\beta)}$. We do not need to consider vertices that are
already covered in both states. Let $cov^{(\alpha)}$ be the subset of
vertices of $G$ that are covered in state ${\vec{x}}^{(\alpha)}$ but
uncovered in state ${\vec{x}}^{(\beta)}$. In the same way we define
$cov^{(\beta)}$. Since all ground states have the same number of
covered vertices both sets must have the same size. Moreover, all
vertices in $cov^{(\alpha)}$ must be neighbors of vertices in
$cov^{(\beta)}$, otherwise the configurations would not be vertex
covers. So the subgraph $G'$ of $G$ which contains all vertices in
$cov^{(\alpha)}$ and $cov^{(\beta)}$ and all the edges from $G$
running between these vertices is a bipartite graph.

\begin{figure}[htbp]
  \centering \includegraphics[width = 8cm]{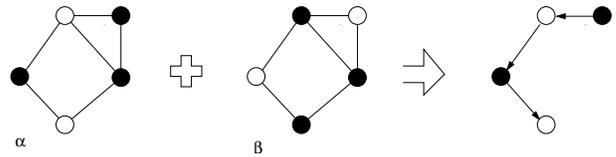}
  \caption{Can one reach state $\beta$ from state $\alpha$ 
just by moving one cover mark at a time, never uncovering any edge?
The algorithm tries to find the answer by looking at the bipartite
graph induced by all nodes that are covered either in $\alpha$ or in
$\beta$ (but not in both) }
  \label{fig:exchangedyn}
\end{figure}

The following algorithm moves cover marks on the graph $G'$ to find
out whether ${\vec{x}}^{(\alpha)}$ and ${\vec{x}}^{(\beta)}$ belong to
the same cluster:

\begin{enumerate}
\item select a vertex $v$ in $G'$ which is covered in state
${\vec{x}}^{(\alpha)}$ and which has exactly one neighbor $w$ in $G'$
(i.e. $w$ is covered in state ${\vec{x}}^{(\beta)}$); if no such $v$
exists: stop
\item \label{item:removepair} remove $v$ and $w$ from $G'$, i.e. set
$G':= G' \setminus \{v\} \setminus \{w\}$
\item go to step (1)
\end{enumerate}

Each pair of vertices taken out in step (2) corresponds to moving a
covering mark along the edge connecting $v$ and $w$. Since $w$ is
always the only uncovered neighbor of $v$ the algorithm only visits
states that are ground states. Note that each covering mark is moved
at most once, for this reason, we call this procedure ``ballistic
search'' \cite{alex-bs}. This method has been already applied to study
the ground-state structure of finite-dimensional spin glasses
\cite{alex-valleys-long}.

If the algorithm stops with $G' = \emptyset$ we have found a path in
configuration space between states $\alpha$ and $\beta$ that only goes
through ground states and we know for sure that $\alpha$ and $\beta$
are in the same cluster.

On the other hand, if such a path exists where each covering mark has
to be moved at most once, then the algorithm is guaranteed to find it
\cite{remark2}. We prove this by contradiction. The main reason is
that for two given states $\alpha$ and $\beta$ the cover mark on any
vertex $v$ is moved to the same vertex $w$ in all possible paths,
i.\,e the individual moves are unique, only the order in which they
are done can differ between paths.

Suppose the opposite would be true, i.e. there exists a path $P$ in
which the mark on vertex $v$ is moved to vertex $w$ and a path $P'$ in
which it is moved to vertex $w'$. Take the first vertex $v$ in $P$ for
which this is true. The moves for all cover marks moved prior to $v$
in $P$ are the same in $P'$. So one can do all these moves, afterwards
$v$ has only one uncovered neighbor, namely $w$. Next, move all cover
marks in $P'$, which have not yet been moved. Now $w$ will be covered,
too. But then vertex $v$ and all its neighbors will be covered, which
is impossible in a ground state. Contradiction, there cannot be two
such paths.

Hence, if the algorithm stops in step (a) with $G' \ne \emptyset$,
then no such path exists where each covering mark is moved at most
once.  This means that either ${\vec{x}}^{(\alpha)}$ and
${\vec{x}}^{(\beta)}$ are in different clusters or that they are
connected by a path such that a covering mark has to be moved at least
twice.  To exclude that the clustering is wrong because some
configurations are connected by paths where a clustering mark is moved
more than once, we compare all configurations pairwise with each
other. This means, we use the transitivity of the cluster to exclude
that two configurations are mistaken to be in different clusters
although they are in the same \cite{alex-bs}. And indeed we have
sometimes observed that for three configurations
$\alpha,\beta,\gamma$, paths are found from $\alpha$ to $\beta$ and
from $\beta$ to $\gamma$, but not from $\alpha$ to $\gamma$.

\begin{figure}[htbp]
  \centering \includegraphics[width = 8cm]{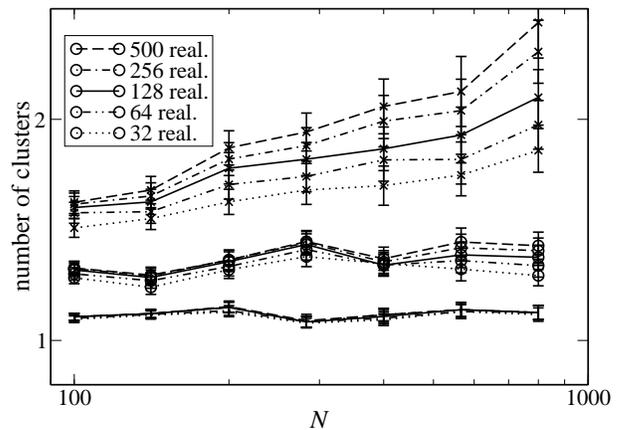}
\caption{Behavior of the number of detected clusters depending on the
number of sampled states for different values of $c$. The crossed
symbols are for $c=4$, circles for $c=3$ and small bars for $c=2$. For
$c<e$ the numbers are confined within error bars, for $c>e$ only a
fraction of the clusters are detected. The number of detected clusters
is still increasing with the number of sampled states. }
\label{fig:neighbourconv}
\end{figure}

For each realization we sample with parallel tempering 500
configurations during a time of $10^6$ MC steps as described in
section \ref{sec:ptalgo}. This number of configurations is far high
enough to ensure that never two configurations belonging actually to
the same cluster are mistaken to be in different clusters. On the
other hand, it might happen that for some (small) clusters no
configurations are sampled. We have tested this explicitly by
calculating the number of clusters as a function of the number of
configurations included in the clustering, see Fig.\
\ref{fig:neighbourconv}. One can see that for small connectivities the
number of clusters is more or less independent on the size of the
sample, while for larger values of $c$ and larger system sizes, the
number of clusters increases slightly with the sample size. This means
that in Fig.\ \ref{fig:clusterneighbour}, where we show the average
number of clusters we find for different connectivities (for the
largest sample size), we have basically a lower bound for the real
average number of clusters. The main result is unaffected by this
sample-size effect: For small connectivities $c$ the number of
clusters is close to one, independent of the system size, and for
large values of $c$ the number of clusters grows.  The results are
compatible with the change appearing near $c=e$, but we cannot
determine the value of the change precisely from our data. Also we
cannot be sure that the growth of the number of clusters is only
logarithmically with system size.  But it seems likely that the number
of clusters grows slower than exponentially with system size, since
for $c=4, N=800$ we find on average less than three clusters. Hence,
this is different from the 1-RSB phase of the satisfiability problem
\cite{BiMoWe,MeZePa,zecchina2004}.  This slow growth is compatible
with the analytical result that for $c>e$ the 1-RSB solution is not
the correct one \cite{zhou2003}, hence a higher level of RSB is to be
expected.

\subsection{Extensive eigenvalues and the number of clusters}

In this section we will use a completely different method to analyze
the structure of the solution space. Sinova et
al. \cite{Sinova2000,Sinova2001} describe a tool for counting
independent pure states in Ising spin glasses.  Here we summarize the
basic aspects of their method. Their main idea is to study the
spectral properties of the spin-spin correlation matrix $\langle S_i
S_j\rangle \equiv C_{ij}$ where $\langle\ \rangle$ indicates the
thermal average. This matrix is semidefinite and since $\langle S_i
S_i\rangle = 1 \forall i$ it has trace $N$. For spin-glasses above the
ordering temperature $T_c$, all eigenvalues are of order one. Below
$T_c$, long-range order appears. If there is a single pair of pure
states, then in the low temperature limit $T\rightarrow 0$,
$C_{ij}\rightarrow \pm 1$, $C$ has one eigenvalue which approaches $N$
as $T\rightarrow 0$, and the rest of the eigenvalues decays to zero
with a power-law in $N$.  So one can detect the presence of long range
order just from analyzing the spectrum of $C_{ij}$.

In the frozen, disordered phase, the phase space breaks up into many
pairs of pure states. They are characterized by their clustering
property \cite{mezard1987}, which we will explain in more detail in
the next subsection \ref{sec:ward}.  Sinova at al. argue that the
number number of extensive eigenvalues of $C_{ij}$ corresponds to the
number of independent pure states of the system. This makes it
possible to detect RSB, which must be present, if the correlation
matrix has more than one extensive eigenvalue. Note that this way of
looking at the phase-space structure is different from looking at the
clusters: The number of clusters may grow exponentially with the
system size, while the number of independent pure states can never be
larger than $N$, since a $N\times N$ matrix has only $N$ eigenvalues.

We apply this method directly to the vertex-cover problem. For every
realization we calculate $C_{ij}$ averaged over the configurations
sampled by parallel tempering with $S_i=1$ if vertex $i$ is covered
and $S_i = -1$ if it is uncovered. We calculate the three largest
eigenvalues and average over $100$ to $400$ realizations, depending on
the system size.

\begin{figure}
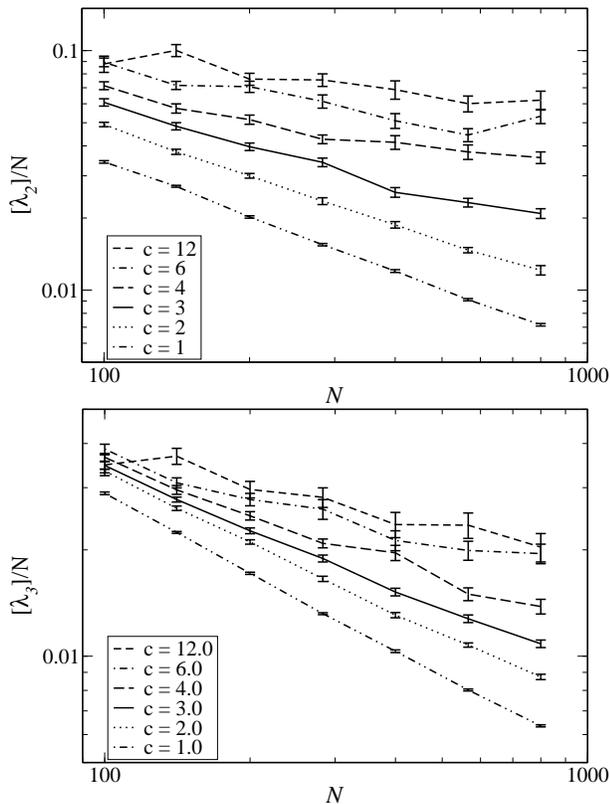

  \centering \includegraphics[width = 8cm,clip]{eigen.eps}
\includegraphics[width = 8cm,clip]{eigen2.eps}
\caption{Scaling of the second largest (top) and third largest
  (bottom) eigenvalue of $C_{ij}$}
  \label{fig:eigenval}
\end{figure}

In Fig.\ \ref{fig:eigenval} we show our results for different values
of $c$ and $N$ at $\mu = 9$. As one can see in the next section, this
$\mu$ is large enough to allow for a nontrivial behavior.  We plot the
normalized value of the second and third largest eigenvalue as a
function of system size.  As expected for $c=1$ and $c=2$ the system
is found to be in the replica symmetric phase: There is only one
extensive eigenvalue, the second and the third decay with a power of
$N$.

For very large $c$ the behavior is different. The second largest
eigenvalue reaches a plateau value around $N=200..300$. The closer the
system is to the $c=e$ the later this plateau is reached. Especially
for $c=3$ the behavior is not yet clear from the reachable system
sizes. The same applies to the third eigenvalue, although one can see
a difference between the largest and the smallest values of $c$.
However, with the reachable system sizes we cannot rule out the
possibility that the third eigenvalue slowly decays for all
connectivities.

Supposing that the behavior of $[\lambda_2]$ does not change again for
large $N$, we conclude that RSB must be present starting from a value
of $2<c<4$ . Please note that we cannot distinguish between 1-RSB and
higher order of RSB from this method. For this reason, we have applied
another method described in the next section.

\subsection{Hierarchical Clustering Approach}

\label{sec:ward}

In this last subsection, we will use a clustering approach that
organizes the states in a hierarchical structure. Such clustering
methods \cite{JainDubes} are widely used in general data analysis,
sometimes also used in statistical mechanics, see e.g. Refs.\
\onlinecite{hed2001,ciliberti2003,hed2004}. The methods all start by
assuming that all states belong to separate clusters. Similarity
between clusters (and states) is defined by a measure called
\emph{proximity matrix} $d_{\alpha,\beta}$.  At each step two very
similar clusters are joined and so a hierarchical tree of clusters is
formed.

A valid hierarchical clustering implies a true ultrametric structure
\cite{RaToVi}. Such a structure is a very important property of the
Parisi-RSB solution \cite{parisi2} of the mean-field SK-model: All
triples $({\vec{x}}^{(\alpha)}, {\vec{x}}^{(\beta)},
{\vec{x}}^{(\delta)})$ of ground states form isosceles triangles with
the third side shorter or equal to the other two sides.

We will try to detect a hierarchical structure in the phase space of
finite-size instances of VC.  As proximity measure for two initial
clusters, each containing only a single state, we naturally choose the
hamming distance between these two states as defined in Sec.\
\ref{sec:hamdist}, divided by the number of vertices. At each step the
two clusters $C_\alpha$ and $C_\beta$ with the minimal distance are
merged to form a new cluster $C_\gamma$. Then the proximity matrix is
updated by deleting the distances involving $C_\alpha$ and $C_\beta$
and adding the distances between $C_\gamma$ and all other clusters
$C_\delta$ in the system. So we need to extend the proximity measure
to clusters with more than one state, based on some suitable update
rule which is usually a function of the distances $d_{\alpha,\beta}$,
$d_{\alpha,\delta}$ and $d_{\beta,\delta}$.

The choice of this function is a widely discussed field since it can
have a great impact on the clustering obtained \cite{JainDubes}. It
should represent the natural organization present in the data and not
some artificial structure induced from the choice of the update rule.
Here we will use \emph{Ward's method} (also called \emph{minimum
variance method}) \cite{ward}. The distance between the merged cluster
$C_\gamma$ and some other cluster $C_\delta$ is given by

\begin{equation}
  \label{eq:warddist}
  d_{\gamma,\delta} =
\frac{(n_\alpha+n_\delta)d_{\alpha,\delta}+(n_\beta+n_\delta)d_{\beta,\delta}
-(n_\alpha+n_\beta)d_{\alpha,\beta} }{n_\alpha+n_\beta+n_\delta}\,
\end{equation}

where $n_\alpha, n_\beta, n_\delta$ are the number of elements in
cluster $C_\alpha,C_\beta,C_\delta$, respectively.  Heuristically
Ward's method seems to outperform other update rules. The choice
guarantees that at each step the two clusters to be merged are chosen
in a way that the variance inside each cluster summed over all
clusters increases by the minimal possible amount.

\begin{figure*}[t]
  \centering
\begin{tabular}{p{5.5cm}p{5.5cm}p{5.5cm}}
\vspace{-\ht\strutbox} \includegraphics[width =
5cm]{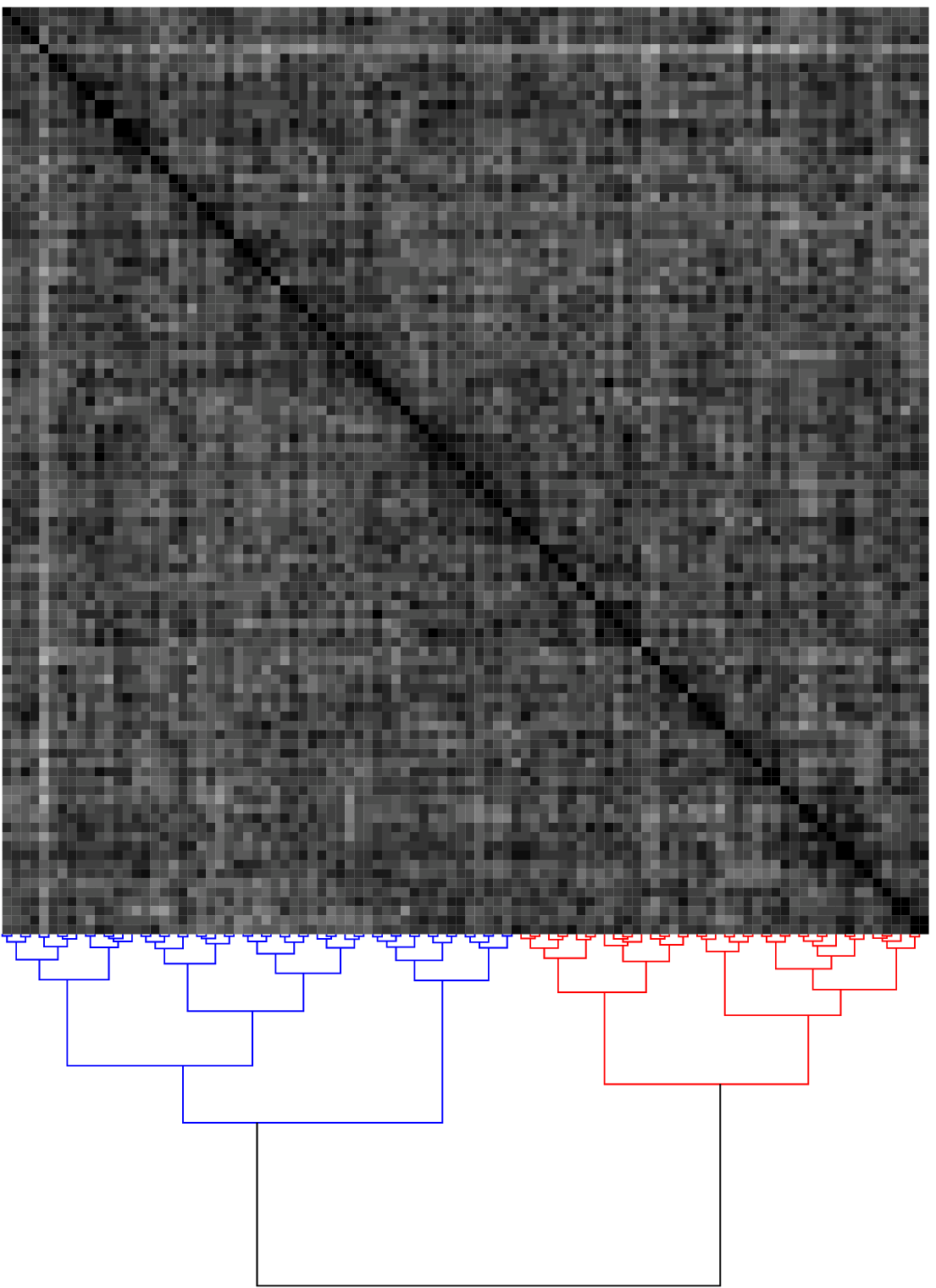} & \vspace{-\ht\strutbox}
\includegraphics[width = 5cm]{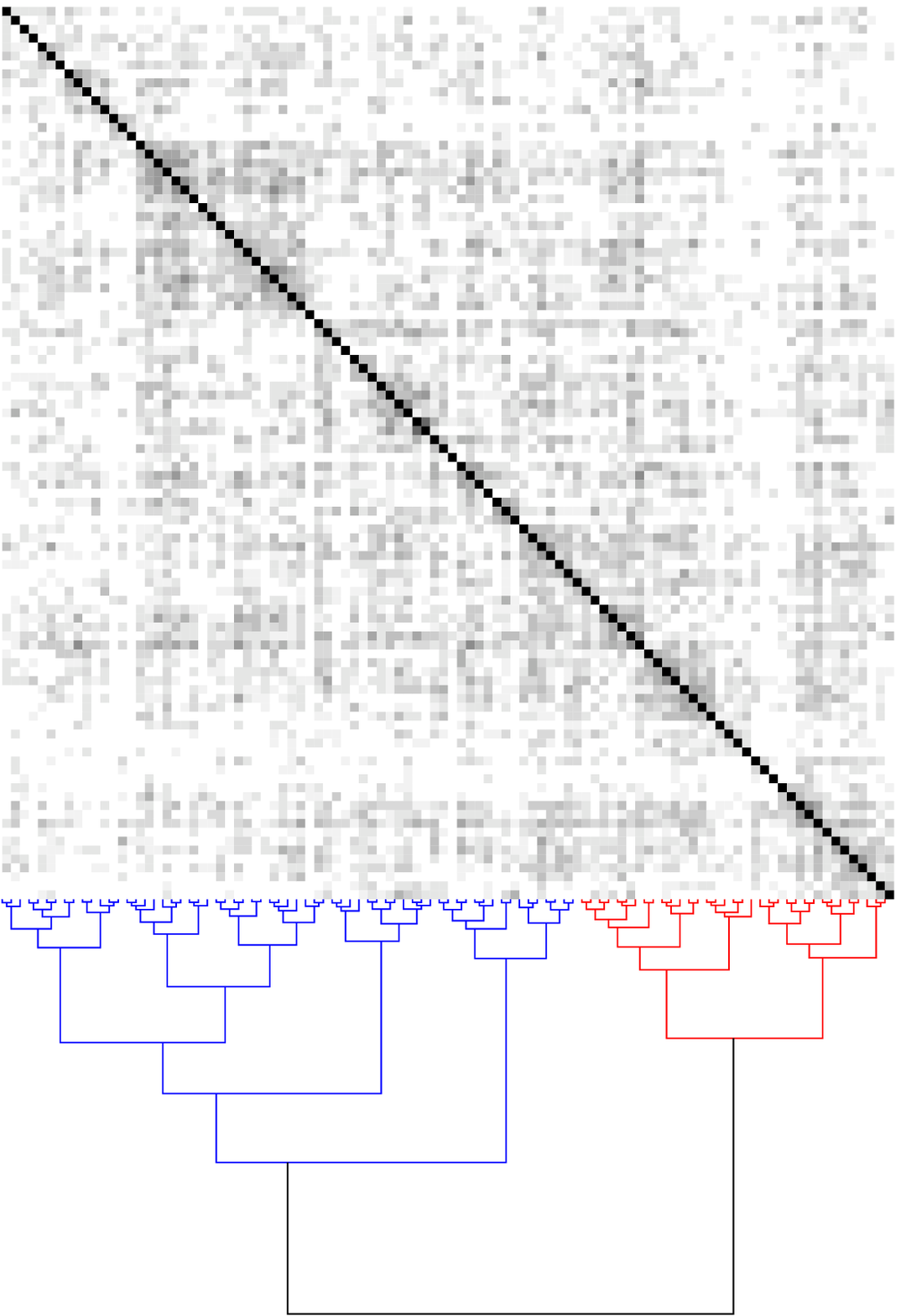} &
\vspace{-\ht\strutbox} \includegraphics[width =
5cm]{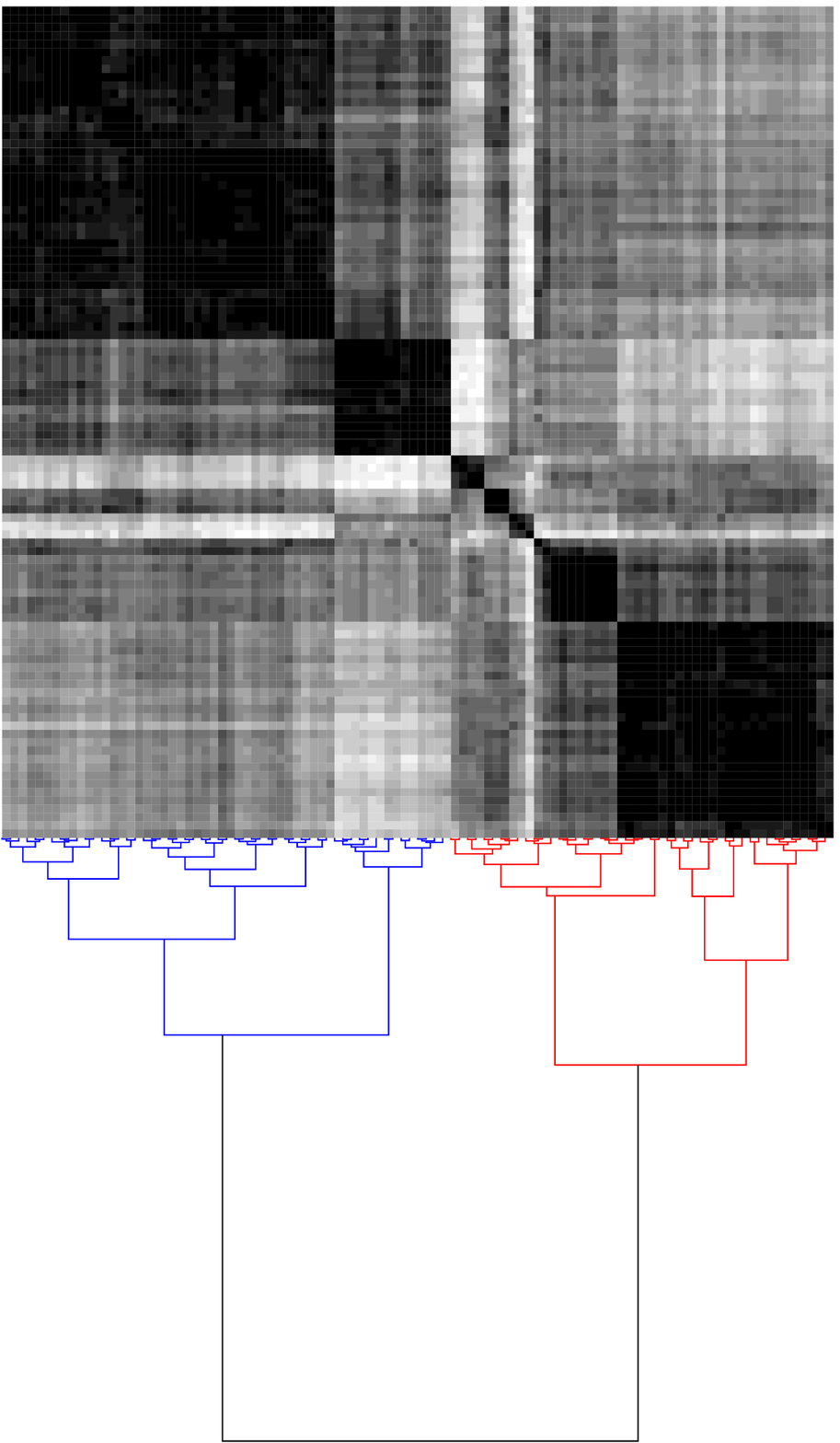}
\end{tabular}
  \caption{Sample dendograms of 100 ground states for a graph with 400
    vertices. Darker colors correspond to closer distances. The left
one is at $c=2$ and $\mu = 9$, i.e. in the RS phase. There is no
structure present. The same is true for $c=6$ and $\mu = 2$. For $c=6$
and $\mu = 9$ the dendogram provides a structure, where the ground
states form clusters. The careful reader may recognize a second or
third level of clustering in this right picture.}
  \label{fig:dendo}
\end{figure*}

The output of the clustering algorithm can be represented as a
\emph{dendogram}. This is a tree with the ground states as leaves and
each node representing one of the clusters at different levels of
hierarchy, see the bottom half of the examples in Fig.\
\ref{fig:dendo}.

Note that Ward's algorithm is able to cluster any data, which can be
always displayed as a dendogram, even if no structure is presented.
Hence, one has to perform additional checks.  A visual check is to
plot the hamming distances as a matrix where the rows and columns are
ordered according to the dendogram. This is shown in the top half of
Fig.\ \ref{fig:dendo}. Darker colors correspond to smaller distances.
The figure shows three different realizations: For small values of
$\mu$ no cluster structure is present. For small values of $c<e$ and
large values of $\mu$, the system is in the RS phase, only a single
cluster is present. For larger values of $c$ and high values of $\mu$,
the ordering of the states obtained by the clustering algorithm
reveals an underlying structure which can be seen in the right part of
the figure. One can see that the states form groups where the hamming
distance between the members is small (dark colors) while the distance
to other states is large. Thus, our results are compatible with
clustering being present for realizations with $c>e$. If you look
carefully you can see more structure inside the clusters. Multiple
levels of clustering indicate higher levels of RSB which we expect to
be present for these values of $c$ \cite{WeHa2,zhou2003}.

To check more quantitatively whether the cluster structure detected by
the algorithm is actually present in the data we evaluate the
\emph{cophenetic correlation coefficient}
\begin{equation}
{\cal K}\equiv [ d\cdot d_c]_G - [ d ] [ d_c ]_G\,,
\end{equation}
where $[\ldots]_G$ denotes the average over the disorder.  This
coefficient measures the correlation between the original distance $d$
of two states and their \emph{cophenetic distance} $d_c$ imposed by
the clustering.  $d_c$ is measured on the dendogram as the distance
given by Eq.  \eqref{eq:warddist} of the two largest clusters that
contain only one of the states.

The results of this test are shown in Fig.\ \ref{fig:dendo}. The
averages are over all samples generated with parallel tempering
(cf. Sec.\ \ref{sec:ptalgo}). As one sees, there is no correlation for
small values of $c$. This is as expected, because for $c<e$ no cluster
structure is present. $\cal K$ increases with increasing magnitude of
the connectivity. In particular, the different curves for $N>100$
cross near $c=e$. For small values of $c$, $\cal K$ decreases with
growing system size, while for large values of $c$, $\cal K$
increases.  This indicates again that around $c=e$ a hierarchical
organization of the VC solution space sets in.  However, for larger
values of $c$ the average correlation seems to converge to a value
close to ${\cal K} \approx 0.8$.  This means that the clustering
imposed by Ward's method does not fully represent the structure
inherent to minimal VCs.

\begin{figure}
  \centering \includegraphics[width = 8cm]{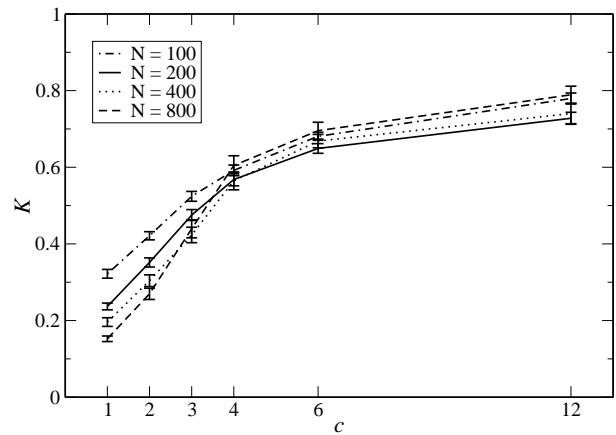}
  \caption{The correlation between hamming distance and cophenetic
    distance measured on the dendogram increases with $c$}

  \label{fig:dendocorrel}
\end{figure}

\section{conclusion}
In our paper we analyzed the ground-state properties of the
vertex-cover problem. Especially we focussed on the phenomenon of
clustering. We found that for connectivities $c<e$ basically only one
ground-state cluster is present. For larger connectivities the number
of numerically detected clusters increases, apparently
logarithmically.  This is compatible with the fact that in analytical
calculations, for $c>e$, the replica-symmetric solution is not longer
valid and the level of RSB seems to be higher than 1-RSB. More
evidence for the appearance of RSB was found by analyzing the spectral
properties of the vertex-vertex correlation function: For $c>e$ two or
more eigenvalues are extensive which can only be the case, if RSB is
present.

With a clustering approach using Ward's algorithm, we tried to detect
directly a hierarchical structure in the ground states. We find
qualitatively higher levels of clustering present in the ground-state
structure for high values of $c$. This would indicate higher level of
replica symmetry breaking.  Also, for $c>e$, the clustering imposed by
the algorithms becomes more and more compatible with the structuring
of the state space.

In summary, the different algorithms are able to find indications for
RSB in the solution landscape of combinatorial optimization problems.
Note that the presence of RSB does not necessarily mean that it is the
same type of RSB, which is found in the solution of the SK model. The
details of the organization of the solution space, e.g. the extent of
ultrametricity, can be different. This can be seen in the convergence
of the cophenetic correlation coefficient to a value apparently
smaller than one.

From our results, which support the previous analytical findings, we
conclude it seems promising to apply the methods to other more
complicated ensembles of VC or to other optimization problems, where
less analytical results are available, in order to understand their
behavior better.

\begin{acknowledgments}
The authors obtained financial support from the {\em
VolkswagenStiftung} (Germany) within the program ``Nachwuchsgruppen an
Universit\"aten''.  We thank M. Weigt for countless hours of fruitful
discussions.  AKH thanks G. Woeginger for interesting discussions at
the Dagstuhl seminar 01091 ``Algorithmic Techniques in Physics''.

\end{acknowledgments}

\appendix
\section{The solution space of vertex cover on a tree consists only of
a single cluster}
\label{app:prooftree}
In section \ref{sec:hamdist} we defined a cluster $C$ as a maximal set
of ground states such that for each pair ${\vec{x}}^{(\alpha)},
{\vec{x}}^{(\beta)} \in C$ there exists a series
$({\vec{x}}^{(\delta_i)})_{i=0 \dots k}$ of ground states with
${\vec{x}}^{(\delta_0)} = {\vec{x}}^{(\alpha)}$ and
${\vec{x}}^{(\delta_k)} = {\vec{x}}^{(\beta)}$ and
$dist_{ham}({\vec{x}}^{(\delta_l)}, {\vec{x}}^{(\delta_{l+1})}) = 2$,
i.e. minimal hamming distance between consecutive elements of the
series. In this appendix we will show, that for trees there can be
only one cluster. The proof will be by induction on the number $N$ of
vertices in the tree.

For $N=2$ there are two ground states, which have hamming distance 2.

\begin{figure}[htbp]
  \centering \includegraphics[width = 8cm]{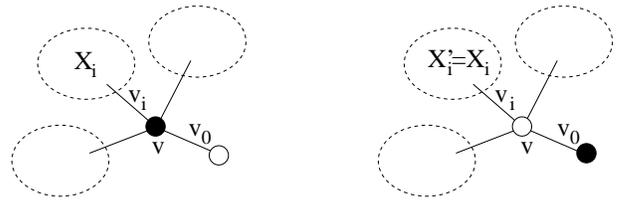}
  \caption{If there is a ground state with vertex $v$ uncovered, then
    all subgraphs must have the same number of covered vertices in
both states.}
  \label{fig:nocl}
\end{figure}

Suppose we have proven the statement for $N$ and consider a graph of
size $N+1$. First note that there is at least one ground state
${\vec{x}}^{(\delta)}$ with a vertex $v$ covered that is a neighbor of
a leaf $v_0$. Such a ground state can be constructed e.g. using leaf
removal. We show separately, that ${\vec{x}}^{(\delta)}$ is in the
same cluster as all ground states $\{{\vec{x}}^{(\delta')}\}$ with $v$
covered and with $v$ uncovered.

Let ${\vec{x}}^{(\delta')}$ be a ground state with $v$ covered. If we
delete vertex $v$ from the tree, then it falls apart into components
$G_1, \dots G_k$ where $k$ is the connectivity of
$v$. ${\vec{x}}^{(\delta)}$ induces a cover on each $G_i$ which is
also a minimal cover on each subgraph, since we started with a minimal
cover on $G$. The same is true for ${\vec{x}}^{(\delta')}$. Each of
the subgraphs has size smaller than $N$, so by induction we can
construct a series from ${\vec{x}}^{(\delta)}$ to
${\vec{x}}^{(\delta')}$ separately on each subgraph, hence both ground
states are in the same cluster.

Now consider a ground state ${\vec{x}}^{(\delta')}$ that has $v$
uncovered. Again we consider the subgraphs one gets by removing $v$
from the graph. Let $X_i$ be the number of covered vertices in the
cover induced from ${\vec{x}}^{(\delta)}$ on the subgraph $G_i$,
analogues let ${X'}_i$ be the number of covered vertices in the cover
induced from ${\vec{x}}^{(\delta')}$. Since ${\vec{x}}^{(\delta)}$ and
${\vec{x}}^{(\delta')}$ both are ground states we have $\sum_i X_i +
1= \sum_i {X'}_i$ which is equivalent to $1 = \sum_i({X'}_i -
X_i)$. All summands on the right side must be non negative, otherwise
${\vec{x}}^{(\delta)}$ would not be a ground state. So there exists
exactly one subgraph $G_j$ with ${X'}_i - X_i = \delta_{i,j}$. This
subgraph must be the leaf $v_0$. For $i\ne j$ the covers induced by
${\vec{x}}^{(\delta')}$ on $G_i$ must be ground states of the
subgraph, since ${X'}_i = X_i$.  So by induction we can construct a
series from ${\vec{x}}^{(\delta)}$ to ${\vec{x}}^{(\delta')}$, again
separately on each subgraph $G_i$ for $i\ne j$ and on the subgraph
$\{v\} \cup \{v_0\}$, hence both ground states are in the same
cluster.

Together we showed that all ground states are in the same cluster as
${\vec{x}}^{(\delta)}$, thus there can only be a single cluster of
ground states.


\begin{thebibliography}{99}
\bibitem{reviewSG} Reviews on spin glasses can be found in: K. Binder
and A.P. Young, Rev. Mod. Phys. {\bf 58}, 801 (1986); M. Mezard,
G. Parisi, M.A. Virasoro, {\it Spin glass theory and beyond}\/, (World
Scientific, Singapore 1987); K.H. Fisher and J.A. Hertz, {\em Spin
Glasses}\/, (Cambridge University Press, Cambridge 1991); A.P. Young
(ed.), {\em Spin glasses and random fields}\/, (World Scientific,
Singapore 1998).

\bibitem{SK} D. Sherrington and S. Kirkpatrick, Phys. Rev. Lett. {\bf
35}, 1792 (1975).

\bibitem{parisi2} G. Parisi, Phys. Rev. Lett. {\bf 43}, 1754 (1979);
J. Phys. A {\bf 13}, 1101 (1980); {\bf 13}, 1887 (1980); {\bf 13},
L115 (1980); Phys. Rev. Lett. {\bf 50}, 1946 (1983).

\bibitem{mezard1987} M. Mézard, G. Parisi, M.A. Virasoro, {\it Spin
glass theory and beyond}\/, (World Scientific, Singapore 1987).

\bibitem{fisher1991} K.H. Fisher and J.A. Hertz, {\em Spin Glasses}\/,
(Cambridge University Press, Cambridge 1991).

\bibitem{RaToVi} R. Rammal, G. Toulouse, and M.A. Virasoro,
Rev. Mod. Phys. {\bf 58} , 765 (1986).


\bibitem{talagrand2003} M. Talagrand, C.R.A.S. {\bf 337}, 111 (2003)

\bibitem{young1983} A.P. Young, Phys. Rev. Lett. {\bf 51}, 1206 (1983)

\bibitem{parisi1993} G. Parisi, F. Ritort and F. Slanina, J. Phys. A
{\bf 26}, 3775 (1993)
 
\bibitem{billore2003} A. Billoire, S. Franz, and E. Marinari,
J. Phys. A {\bf 36} (2003)

\bibitem{Sinova2000} J. Sinova, G. Canright, and A.H. MacDonald
Phys. Rev. Lett. {\bf 85}, 2609 (2000).

\bibitem{Sinova2001} J. Sinova, G. Canright, H.E. Castillo, and
A.H. MacDonald, Phys. Rev. B {\bf 63}, 104427 (2001).

\bibitem{hed2004} G. Hed, A.P. Young, and E. Domany
Phys. Rev. Lett. {\bf 92}, 157201 (2004)

\bibitem{KM00} F. Krzakala and O. C. Martin,
Phys. Rev. Lett. {\bf 85}, 3013 (2000).

\bibitem{palassini2000} M. Palassini and A.P. Young, Phys
Rev. Lett. {\bf 85}, 3017 (2000)


\bibitem{BiMoWe} G. Biroli, R. Monasson, and M. Weigt, Eur. Phys. J. B
{\bf 14}, 551 (2000).

\bibitem{MeZePa} M. M\'ezard, G. Parisi, and R. Zecchina, Science {\bf
297}, 812 (2002).

\bibitem{MeZe} M. M\'ezard and R. Zecchina, Phys. Rev. E {\bf 66},
056126 (2002).

\bibitem{Me1} S. Mertens, Phys. Rev. Lett. {\bf 81}, 4281 (1998).

\bibitem{Me2} S. Mertens, Phys. Rev. Lett. {\bf 84}, 1347 (2000).

\bibitem{MuPaWeZe} R. Mulet, A. Pagnani, M. Weigt, and R. Zecchina,
Phys. Rev. Lett. {\bf 89}, 268701 (2002).

\bibitem{WeHa1} M. Weigt and A.K. Hartmann, Phys. Rev. Lett. {\bf 84},
6118 (2000).

\bibitem{HaWe} A.K. Hartmann and M. Weigt, Theor. Comp. Sci. {\bf
265}, 199 (2001).

\bibitem{WeHa2} M. Weigt and A.K. Hartmann, Phys. Rev. E {\bf 63},
056127 (2001).

\bibitem{zhou2003} H. Zhou, Eur. Phys. J. B {\bf 32}, 265 (2003).


\bibitem{cover-review} A.K. Hartmann and M. Weigt, J. Phys. A {\bf
36}, 11069 (2003).

\bibitem{GaJo} M.R. Garey and D.S. Johnson, \textit{Computers and
intractability} (Freeman, New York, 1979).


\bibitem{review1} Hogg T, Huberman B A and Williams C (eds.), {\it
Frontiers in problem solving: phase transitions and complexity}, {
Artif. Intell.} {\bf 81} (I+II) (1996).

\bibitem{review2} O. Dubois, R. Monasson, B. Selman and R. Zecchina
(eds.), special issue of { J. Theor. Comp. Sci.}\/ {\bf 265} (2001).

\bibitem{monasson1996} R. Monasson and R. Zecchina,
Phys. Rev. Lett. {\bf 76}, 3881 (1996);
Phys. Rev. E {\bf 56}, 1357 (1997)
\bibitem{nature} R. Monasson, R. Zecchina, S. Kirkpatrick, B. Selman,
and L. Troyansky, Nature \textbf{400}, 133 (1999).

\bibitem{weigt2004} M. Weigt, in: A.K. Hartmann and H. Rieger (eds.),
{\em New Optimization Algorithms in Physics}, (Wiley-VCH 2004).

\bibitem{zecchina2004} R. Zecchina, in: A.K. Hartmann and H. Rieger
(eds.), {\em New Optimization Algorithms in Physics}, (Wiley-VCH
2004).

\bibitem{mezard2003} M. Mezard, F. Ricci-Tersenghi, and R. Zecchina,
J. Stat. Phys. {\bf 111}, 505 (2003).


\bibitem{remark1} One can also construct problems with only one
GS-cluster but many metastable states, these problems are hard as
well.

\bibitem{mitchell1992} D. Mitchell, B. Selman and H. Levesque, in {\it
Proc. 10th Natl. Conf. Artif. Intell. (AAAI-92)}, 440 (AAAI Press/MIT
Press, Cambridge, Massachusetts, 1992).

\bibitem{selman1994} B. Selman and S. Kirkpatrick, Science {\bf 264},
1297 (1994).


\bibitem{ErRe} P. Erd\"os and A. R\'enyi, {
Publ. Math. Inst. Hung. Acad. Sci.} {\bf 5},17 (1960).

\bibitem{Bo} B. Bollobas, { Random Graphs} (Academic Press, 1985).

\bibitem{bauer2001a} M. Bauer and O. Golinelli, Phys. Rev. Lett. {\bf
86}, 2621 (2001)

\bibitem{bauer2001} M. Bauer and O. Golinelli, Eur. Phys. J. B {\bf
24}, 339 (2001)

\bibitem{tarjan77} R.E. Tarjan and A.E. Trojanowski , { SIAM J. Comp.}
{\bf 6}, 537 (1977)

\bibitem{marinari1992} E. Marinari and G. Parisi, Europhys. Lett.
{\bf 19}, 451 (1992).

\bibitem{hukushima1996} K. Hukushima and K. Nemoto, J. Phys. Soc. Jpn.
{\bf 65}, 1604 (1996).

\bibitem{landau2000} D.P. Landau and K. Binder, {\em A Guide to Monte
Carlo Simulations in Statistical Physics}\/, (Cambridge University
Press, Cambridge 2000).

\bibitem{marinari1998} E. Marinari, in: J. Kert\'sz and Imre Kondor
(eds.), {\em Advances in Computer Simulation}, 50 (Springer Verlag,
Berlin 1998)




\bibitem{alex-valleys-long} A.K. Hartmann, Phys. Rev. E {\bf 63},
016106 (2001).

\bibitem{alex-bs} A.K. Hartmann, J. Phys. A {\bf 33}, 657 (2000).

\bibitem{remark2} This is different compared to the application of
ballistic search for spin glasses \cite{alex-bs}. There it depends
also on the order of the spin flips whether a path in configuration
space is found or not.




\bibitem{JainDubes} A.K. Jain and R.C. Dubes, {\it Algorithms for
Clustering Data}, (Prentice-Hall, Englewood Cliffs, USA, 1988).

\bibitem{hed2001} G. Hed, A.K. Hartmann, D. Stauffer, and E. Domany,
Phys. Rev. Lett., {\bf 86}, 3148 (2001).

\bibitem{ciliberti2003} S. Ciliberti and E. Marinari, preprint
cond-mat/0304273 (2003).


\bibitem{ward} J. Ward, J. of the Am. Stat. Association {\bf 58}, 236
(1963).

\end{thebibliography}
\end{document}